\documentclass[aps,prd,eqsecnum,twocolumn,nofootinbib,preprintnumbers]{revtex4}
\usepackage{amsmath}
\usepackage{amssymb}
\usepackage{amsfonts}
\usepackage{latexsym}
\usepackage{graphicx}% Include figure files
\usepackage{dcolumn}% Align table columns on decimal point
\usepackage{bm}% bold math
\usepackage{empheq}
\usepackage{color}
\usepackage{times}  %%  Only change the main text font to Times New Roman
\newcommand{\eq}[1]{\begin{equation}#1\end{equation}}

\newcommand{\ea}[1]{\begin{equation}\begin{aligned}#1\end{aligned}\end{equation}}

\newcommand{\lrp}[1]{\left( #1 \right)}  %% parentheses
\newcommand{\lrsb}[1]{\left[ #1 \right]}  %% square brackets
\def\rd{\partial}
\def\vx{\bm{x}}%% vector x
\def\vk{\bm{k}}

\def\va{\bm{A}}
\definecolor{gxhighlight}{rgb}{0.85,0.9,1}
\def\dep{\delta\phi}
\def\dop{\dot{\phi}}
\def\bap{\bar{\phi}}

\begin{document}

%%********************  Article Number.Begin  ********************%%

%\preprint{CAS-KITPC/ITP-088}

%%********************  Article Number.End  ********************%%

%%********************  Title.Begin  ********************%%

\title{Can Relic Superhorizon Inhomogeneities be Responsible for Large-Scale CMB Anomalies?}

\author{Xian Gao}
\email{gaoxian@itp.ac.cn} \affiliation{\vspace{0.3cm}
    Key Laboratory of Frontiers in Theoretical Physics,\\
     Institute of Theoretical Physics, Chinese Academy of
    Sciences\\
    No. 55, Zhong-Guan-Cun East Road, Hai-Dian District, Beijing 100080, P.R.China
}

%\author{Miao Li}
%\email{mli@itp.ac.cn} \affiliation{\vspace{0.3cm}
%    Key Laboratory of Frontiers in Theoretical Physics,\\
%     Institute of Theoretical Physics, Chinese Academy of
%    Sciences\\
%    No.55, Zhong-Guan-Cun East Road, Hai-Dian District, Beijing 100080, P.R.China
%}


%\keywords{Quantum Gravity}%Use showkeys class option if keyword
%                              %display desired

%%********************  Title.End  ********************%%

%%********************  Abstract.Begin  ********************%%

\begin{abstract}
    We investigate the effects of the presence of relic classical
    superhorizon inhomogeneities during inflation. This superhorizon
    inhomogeneity appears as a gradient locally and picks out a
    preferred direction. Quantum fluctuations on this
    slightly inhomogeneous background are generally statistical anisotropic. We find a quadrupole modification to the ordinary isotropic spectrum.
    Moreover,
    this deviation from statistical isotropy
     is scale-dependent, with a $\sim -1/k^2$ factor. This result implies that the
    statistical anisotropy mainly appears on large scales, while the spectrum on small scales remains highly isotropic. Moreover, due to this $-1/k^2$ factor, the power on large scales
    is suppressed. Thus, our model can
simultaneously explain the observed anisotropic alignments of the
low-$\ell$ multipoles and their low power.
    \end{abstract}

%%********************  Abstract.End  ********************%%
%%
%%
\maketitle

%\hrule\vspace{-0.3cm}
%\tableofcontents
%\vspace{0.5cm} \hrule \vspace{0.5cm}
%%
%%********************  Main Article.Begin  ********************%%
%%
%%
\section{Introduction}

The current observation data support the standard $\Lambda$CDM model
\cite{Komatsu:2008hk}. Recently, however, there is an growing
interest in analyzing possible large-scale anomalies of CMB, from
both theoretical and observational sides
\cite{Pullen:2007tu},\cite{Contaldi:2003zv},\cite{Ackerman:2007nb},\cite{Gordon:2005ai},\cite{Land:2005ad},\cite{Erickcek:2008sm},\cite{ArmendarizPicon:2008yr},\cite{ArmendarizPicon:2007nr},\cite{Donoghue:2007ze}\cite{Yokoyama:2008xw}.

It appears that the lowest CMB multipoles are anomalous in two
seemingly distinct aspects. Firstly, the angular power $C_{\ell}$ at
the lowest $\ell$ is abnormally suppressed \cite{Contaldi:2003zv}.
Secondly, they have an improbable directionality revealed by the
fact that for a certain preferred orientation, one $m$-mode absorbs
most of the power, which may imply the presence of an ``axis of
evil" \cite{Land:2005ad,Ackerman:2007nb}.

The statistical properties of perturbations carry the same
symmetries as the background on which they are generated. In the
standard scenario, quantum fluctuations are assumed to be generated
on a spatially homogeneous and isotropic background. Background
spatial inhomogeneities are neglected in most of the analysis.
Indeed, a long period of inflation pulls all non-smooth classical
initial conditions out of the horizon. However, if inflation lasts
the minimal number 60 e-folds or so, perturbations with comoving
wave-numbers of cosmological interest cross the horizon just during
the earliest several e-folds of inflation. Thus, the relic
inhomogeneities may leave some marks in the primordial quantum
fluctuations.

In this Letter, we investigate the possible effects of a relic
classical superhorizon inhomogeneity during inflation, especially
its effects on the statistical properties of quantum fluctuations.
In \cite{Gordon:2005ai,Erickcek:2008sm}, a single superhorizon
perturbation mode were considered to explain the power asymmetry of
CMB on large scales. While in this work, we consider relic
superhorizon inhomogeneities as \emph{background}. Perturbation
theory in the presence of a spatially inhomogeneous inflaton
background value has been investigated by several authors before
\cite{ArmendarizPicon:2007nr},\cite{Donoghue:2007ze}. Quantum
fluctuations on this slightly inhomogeneous background are generally
statistical anisotropic.

The deviations from statistical isotropy can take on many forms,
which may correspond to different physical origins. One simple form
was presented in \cite{Ackerman:2007nb},% which parameterize the
%primordial power spectrum $P(\vk)$ with dependence on the direction
%$\hat{\vk}$ of comoving wave-vector $\vk$ of the form,
$
    P(\vk) = P(k) \lrsb{ 1+ g(k)(\hat{\vk} \cdot \bm{n})^2 }
$.
%In \cite{Ackerman:2007nb}, $g(k)$ was assumed to be approximately
%constant $g_{\ast}$.
In this work, we find that the (leading-order) correction to the
statistically isotropic power spectrum is of the ACW form
\cite{Ackerman:2007nb}, but with a $k$-dependent factor $g(k) \sim
-1/k^2$. Thus, the power spectrum deviates from isotropy mainly on
large scales (small $k$), while remains highly isotropic on small
scales. Moreover, the spectrum itself is also suppressed on large
scales. Thus, our model can simultaneously explain the observed
anisotropic alignments of the low-$\ell$ multipoles and their low
power.

%non-trivial topology \cite{Levin:2001fg}
%
%multipole vectors \cite{Copi:2003kt}
%
%vector field perturbations \cite{Dimopoulos:2008yv}
%
%multi-field models \cite{Yokoyama:2008xw}
%
%anisotropic inflation models \cite{Gumrukcuoglu:2007bx}
%
%Suppression of power on the lowest multipoles
%\cite{Nicholson:2007by}
%
%backreaction \cite{Mukhanov:1996ak}

\section{Model and Background}

In this work we investigate general single scalar field inflationary
models, with an action of the form $S = \int d^4x\,\sqrt{-g}\lrsb{
R/2 +P(X,\phi)}$, where $X = -\frac{1}{2}(\rd\phi)^2$. Now we
consider a slightly inhomogeneous inflaton background $\bap(t,\vx)$.
Firstly, we assume the deviation from homogeneity is very small,
i.e., if $\bap(t)$ is a spatial average of $\bap(t,\vx)$ (e.g. in
one Hubble volume), then we assume that \eq{
    \left| \frac{\bap(t,\vx)-\bap(t)}{\bap(t)} \right| \ll 1\,.
} Secondly, we assume the inhomogeneities are superhorizon. In other
words, the typical comoving scale of these background
inhomogeneities $l$ is much larger than today's comoving Hubble
scale, $l \gg (a_0H_0)^{-1}$. Actually, it has been known long
before that inflation can occur in the presence of superhorizon
initial inhomogeneities \cite{Goldwirth:1991rj}. In other words, the
inflaton field initially should be smooth up to physical scales
larger than $\sim H^{-1}$. Thus, today's observational universe can
indeed inflate from an initial small patch with classical
inhomogeneities with typical comoving scale $l \gg (a_0H_0)^{-1}$.

Superhorizon inhomogeneities locally look like a gradient, $
    \bap(t,\vx) = \bap(t,\vx_0) + (\vx-\vx_0)\cdot
    \nabla \bap(t,\vx_0) +\cdots
$. In this work, we neglect higher-order derivatives and treat
$\nabla \bap$ as approximately constant (while $\bap$ itself is
indeed slow-rolling, for instance).

\section{Linear Perturbations}

We split $\phi(t,\vx)$ into background and fluctuation
configurations, \eq{\label{decomposition}
    \phi(t,\vx) = \bap(t,\vx) + \dep(t,\vx)\;.
    }
In the following we denote $\partial_i \bap(t,\vx) = A_i$, which we
have assumed to be constant.

As a first-step investigation, we neglect the metric perturbations.
Thus the calculation is straight forward since all we have to do is
to expand the action for the scalar field directly. According to
(\ref{decomposition}), simple Taylor expansion of $P(X,\phi)$ around
the background $\bap$ up to second order of $\dep$ gives $
    -\frac{1}{2} \Sigma^{\mu\nu} \rd_{\mu}\dep \rd_{\nu}\dep
    -P_{,X\phi}\rd^{\mu}\bap \rd_{\mu}\dep\,\dep +
    \frac{1}{2}P_{,\phi\phi}\dep^2 \,,
$ where we have defined \eq{
    \Sigma^{\mu\nu} \equiv P_{,X}g^{\mu\nu} - P_{,XX}\rd^{\mu}\bap
    \rd^{\nu}\bap \,.
}
%with \ea{
%    \Sigma^{00} &= - \lrp{ P_{,X} + 2XP_{,XX} } - \frac{P_{,XX}}{a^2}\lrp{ \rd_i\bap}^2 \,,\\
%    \Sigma^{ij} &= \frac{P_{,X}}{a^2} \lrp{ \delta_{ij} -
%    \frac{P_{,XX}}{P_{,X}a^2}\rd_i\bap \rd_j\bap } \,,\\
%    \Sigma^{0i} &= \frac{P_{,XX}}{a^2}\dot{\bap}\rd_i\bap \;.
%}
Note that due to the non-vanishing background gradient
$\rd_{i}\bap$, $\Sigma^{ij}$ is not proportional $\delta_{ij}$ any
more. The presence of the additional term which is proportional
$\rd_i\bap\rd_j\bap$ in $\Sigma^{ij}$ breaks spatial rotational
invariance and will be responsible for the generation of statistical
anisotropies of the scalar perturbations $\dep$.

After introducing a new variable $u(\eta,\vx) =
a\sqrt{-\Sigma^{00}}\dep$, the second-order action for the scalar
field perturbations can be written as \ea{{\label{action}}
    S &= \int d\eta \frac{d^3k}{(2\pi)^3} \frac{1}{2} \left[ u'_{-\vk}u'_{\vk} + \frac{2i\sqrt{\epsilon}\gamma}{aH}(\va\cdot\vk)u'_{-\vk}u_{\vk}
    \right.\\
    &\left.\qquad - \lrp{ c^2k^2 -\frac{a''}{a} + \mathcal{M} + \frac{\gamma(\va\cdot\vk)^2}{a^2H^2} }
    u_{-\vk}u_{\vk} \right] \,,
} where $\eta$ is comoving time, and a prime represents $\rd/\rd
\eta$, $\mathcal{H}\equiv a'/a$, and $\mathcal{M}$ is an effective
mass term, $\epsilon$ is defined as $\epsilon = \dop^2/H^2$. Here we
define two dimensionless parameters \eq{
    c^2 = \frac{P_{,X}}{-\Sigma^{00}} \,,\qquad
    \gamma = \frac{H^2P_{,XX}}{\Sigma^{00}} \,.
}  In this Later, we assume $\gamma>0$. Note that the canonical case
corresponds to $c^2=1$ and $\gamma=0$. In deriving (\ref{action}),
we use the approximation that $\Sigma^{\mu\nu}$ is a function of
only $\eta$, that is, we neglect the spatial dependence of
$\Sigma^{\mu\nu}$.

\subsection{Equation of Motion for the Perturbations and Solutions}

The classical equation of motion for the mode function according to
the second-order action (\ref{action}) is \ea{{\label{mode_eom}}
    & u''_{\vk} + \frac{2i\sqrt{\epsilon}\gamma (\va\cdot\vk)}{aH}  u'_{\vk}\\
    &\qquad  + \lrp{ c^2 k^2 - \frac{a''}{a} + \mathcal{M} a^2 + \frac{\gamma (\va\cdot \vk)^2}{a^2H^2} } u_{\vk}=
    0 \,.
} As it can be seen, an anisotropic dispersion relation arises,
which will be responsible for the generation of a statistically
anisotropic power spectrum.

Now we take the scale factor as $a(\eta) = -1/H\eta$. Remarkably, in
this simplest case, Eq. (\ref{mode_eom}) has an analytic
solution, \eq{{\label{mode_solution}}
    u_{\vk}(\eta) =
    \frac{\Gamma(\alpha-\nu) }{2^{\nu+1}\sqrt{\pi}} e^{\frac{i\pi}{2}(\nu+\frac{1}{2})}  e^{\frac{i}{2}\lambda \eta^2 }\sqrt{-\eta}(-c
    k\eta)^{\nu}U(\alpha,\nu+1,z) \,,
} in which $U(\alpha,\nu+1,z)$ is the confluent hypergeometric
function, with \ea{{\label{parameters}}
    \nu &= \sqrt{\frac{9}{4} - \frac{\mathcal{M}}{H^2}} \,,\\
    \alpha &= \frac{1}{2}(\nu+1) - \frac{ic^2k^2 - \sqrt{\epsilon}\gamma (\va\cdot\vk)}{4|\va\cdot\vk| \sqrt{\gamma +\epsilon \gamma^2}}\,,\\
    z &= -i |\va\cdot\vk| \sqrt{\gamma +\epsilon \gamma^2} \eta^2\,,\\
    \lambda &= |\va\cdot\vk|\sqrt{\gamma +\epsilon \gamma^2} + \sqrt{\epsilon} \gamma
    (\va\cdot\vk) \,.
} Due to the presence of the factor $\va\cdot\vk$, the modes
generally depend on $\vk$ rather than $k=|\vk|$. Here the
coefficient in (\ref{mode_solution}) is chosen in order to satisfy
the Wronskian normalization condition
$u_{\vk}'(\eta)u_{\vk}^{\ast}(\eta)-
u_{\vk}'^{\ast}(\eta)u_{\vk}(\eta)= i$. In the limit $A_i\rightarrow
0$, it can be verified that (\ref{mode_solution}) reduces to the
well-known functional form $
    u_{\vk}(\eta) \xrightarrow[]{A_i\rightarrow 0}
    \frac{\sqrt{\pi}}{2}e^{\frac{i\pi}{2}\lrp{\nu+\frac{1}{2}}}\sqrt{-\eta}
    H^{(1)}_{\nu}(-ck\eta) \,,
$ which describes nothing but the normalized mode function of a
massive scalar field in pure de Sitter spacetime. Thus our mode
solution (\ref{mode_solution}) generalizes the standard homogeneous
and isotropic background to the case of the presence of superhorizon
background inhomogeneities.

\subsection{Anisotropic Power Spectrum}

\begin{figure}[h]
    \centering
    \begin{minipage}{0.45\textwidth}
    \includegraphics[width=7cm]{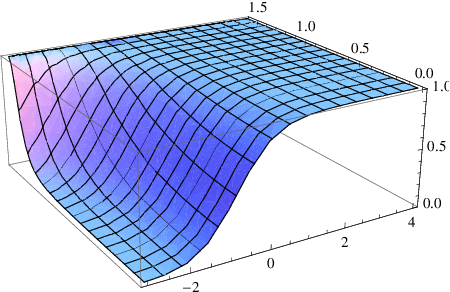}
    \caption{Deformation factor $|\mathcal{A}(\vk)|^2$ as function of $k=|\vk|$ and $\theta$, where $\cos\theta \equiv \hat{\va}\cdot\hat{\vk}$. Parameters are chosen as $c=2$, $A=\gamma=1$, $\epsilon=0.01$, $\nu=1.52$. We plot $\ln k$ from -3 to 4, and $\theta$ from 0 to $\pi/2$.}
    \label{fig_1}
    \end{minipage}
\end{figure}

When all modes of cosmological interest exit the Hubble scale, that
is, in $\eta\rightarrow 0$ limit, since $U(\alpha,\beta,z)
\rightarrow \frac{\Gamma(\beta-1)}{\Gamma(\alpha)} z^{1-\beta}$ as
$z\rightarrow 0$ (when $\beta>2$), we get
\eq{{\label{mode_sol_superhorizon}}
    u_{\vk}(\eta) =
    \mathcal{A}(\vk)\,
    e^{\frac{i\pi}{2}(\nu-\frac{1}{2})}2^{\nu-\frac{3}{2}}\frac{\Gamma(\nu)}{\Gamma(3/2)}\frac{1}{\sqrt{2ck}}\lrp{-ck\eta}^{\frac{1}{2}-\nu}
    \,,
} where we have defined an anisotropic deformation factor,
\eq{{\label{deformatio_factor}}
    \mathcal{A}(\vk) \equiv \frac{\Gamma(\alpha-\nu)}{\Gamma(\alpha)}\lrp{\frac{4i |\va\cdot\vk|\sqrt{\gamma+\epsilon \gamma^2}}{c^2k^2}
    }^{-\nu} \,,
} which is responsible for the anisotropic deformation of the power
spectrum on large scales. Here $\alpha,\nu$ are given in
(\ref{parameters}). In the standard scenarios, $\mathcal{A}(\vk)$ is
just $\mathcal{A}(\vk) = 1$. In our case, $\mathcal{A}$ depends on
$\vk$, not only on its amplitude $k$, but also on it direction, more
precisely, on $\hat{\va}\cdot\hat{\vk}$ (see Fig. \ref{fig_1}).

We would like to investigate the leading order effect of $A_i =
\rd_i\bap$. Taking the limit $A_i \rightarrow 0$ in
(\ref{deformatio_factor}) and using  Stirling's formula, we get
\ea{{\label{df_leading}}
    \mathcal{A}(\vk) &= 1 - i\frac{ \nu \sqrt{\epsilon} \gamma (\va\cdot\vk)}{c^2k^2} \\
    &\quad- \frac{ \gamma (\va\cdot\vk)^2}{c^4k^4} \lrsb{ \frac{2}{3}\nu(\nu^2-1) + \frac{\nu(\nu+1)(4\nu-1)}{6}\epsilon \gamma
    } \,.
} It can be seen directly from (\ref{df_leading}) that the
anisotropic deformation factor $\mathcal{A}(\vk)$ reduces to 1 when
setting $A_i=0$, as expected. Moreover, for those modes with
wavenumbers $\vk$ perpendicular to $\va$, i.e. $\va\cdot\vk=0$, it
will be shown that $\mathcal{A}(\vk) = 1$, that is, these modes do
not feel the presence of $\va$, and thus get no corrections. Modes
with $\va\cdot\vk \neq 0$ will get corrections from $\va$, but in
the limit $\va\rightarrow 0$ (or more precisely, typical scale of
$k$ much larger than $A$, $k \gg |\va|$), it will also be shown that
$\mathcal{A}(\vk)\rightarrow 1$, which means that small scales get
smaller anisotropic corrections from $\va$.

\begin{figure}[h]
    \centering
    \begin{minipage}{0.4\textwidth}
    \includegraphics[width=7cm]{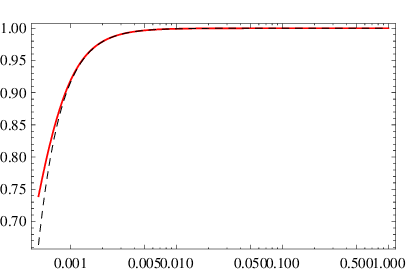}
    \caption{Leading-order quadrupole terms of $|\mathcal{A}(\vk)|$ (for a given direction of $\hat{\vk}$). The black dashed curve is the exact result, the red curve corresponds to the quadrupole term given in (\ref{df_leading}). Here we have chosen that $\vk$ is parallel to $\va$ without loss of generality. Parameters are chosen as $A=10^{-3}$, $c=2, n_s=0.96, \epsilon=0.01$. We plot $k$ from $0.5 \times 10^{-3}$ to 1.}
    \label{fig_leading}
    \end{minipage}
\end{figure}

From (\ref{df_leading}), \eq{{\label{quadrupole}}
    |\mathcal{A}(\vk)|^2 = 1 -
    \frac{ \gamma(\va\cdot\vk)^2}{c^4k^4} \lrsb{ \frac{4}{3}\nu(\nu^2-1) + \frac{\nu(4\nu^2-1)}{3}\epsilon \gamma } \,.
} It is interesting to note that the leading order correction term
are quadrupole, there is no dipole correction. Thus, our result
gives the anisotropic spectrum of ACW form \cite{Ackerman:2007nb},
$P(\vk) = P(k) \lrsb{1 + g(k) (\hat{n} \cdot \vk)^2}$  with \eq{
    g(k) = -\frac{\gamma A^2}{c^4k^2}\lrsb{ \frac{4}{3}\nu(\nu^2-1) + \frac{\nu(4\nu^2-1)}{3}\epsilon \gamma
    } \,.
} In our model, $g(k)<0$ and it has an apparent $k$-dependence,
$g(k) \sim - 1/k^2$. This result has two implications. First, the
deviation from statistical isotropy of the power spectrum is
\emph{not} constant on all scales (while in \cite{Ackerman:2007nb}
and other works it is often assumed that $g(k)$ is a constant
$g_{\ast}$). Statistical anisotropy of $C_{\ell}$ spectrum mainly
appears at the lowest $\ell$'s, while the spectrum at higher
$\ell$'s remains highly isotropic (this can be seen clearly from
Fig.\ref{fig_1}, where for smaller $k$, $|\mathcal{A}(\vk)|$ is
highly anisotropic, while for larger $k$, $|\mathcal{A}(\vk)|$
remains flat). Second, for a given direction $\hat{\vk}$, since
$g(k) \sim -1/k^2$, the spectrum itself is suppressed at large
scales (see Fig. \ref{fig_leading}). This phenomenon explains the
loss of power at the lowest $\ell$'s in the $C_{\ell}$ spectrum.

Finally, the power spectrum of scalar perturbation on superhorizon
scales reads, \ea{{\label{final_spectrum}}
    \Delta^2_{\dep}(\vk) &\equiv \frac{k^3}{2\pi^2}|\dep_{\vk}|^2 \\
    &\simeq \lrp{ \frac{H}{2\pi} }^2\lrp{ \frac{ck}{aH}}^{3-2\nu}
    \frac{|\mathcal{A}(\vk)|^2}{cP_{,X}} \\
    &= \Delta^2_{\dep}(k)\, |\mathcal{A}(\vk)|^2\,,
} where $|\mathcal{A}(\vk)|^2$ is given in (\ref{quadrupole}). The
general anisotropic power spectrum (\ref{final_spectrum}) is
function of $k$ and $\theta$. In our model, its \emph{shape} is
similar to that of $|\mathcal{A}(\vk)|^2$ (see Fig. \ref{fig_1}).
%Note that in $\mathcal{A}(\vk) \rightarrow 1$ limit, our anisotropic
%spectrum (\ref{final_spectrum}) reduces to the well-known result in
%general single field inflation models.

\subsection{Suppression of Power on the Largest Scales}

\begin{figure}[h]
    \centering
        \begin{minipage}{0.4\textwidth}
        \includegraphics[width=7.7cm]{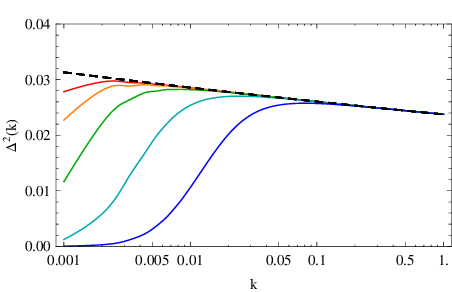}
    \caption{Suppression of the power on the largest scales (for a given direction $\hat{\vk}$). The dashed line is traditional isotropic spectrum with $n_s=0.96$. From top to bottom, $A= 0.3\times 10^{-3}, 0.5\times 10^{-3}, 10^{-3}, 0.3\times 10^{-2}, 10^{-2}$. Parameters are chosen as $\epsilon=0.01$, $\gamma=H=1$, $c=1.2$. The range of $k$ is from $10^{-3}$ to 1.}
    \label{fig_suppresion}
    \end{minipage}
\end{figure}

The observational data appear to suggest a spatial modulation in the
CMB spectrum. Especially, a vanishing CMB temperature
auto-correlation on the largest scales has been investigated by
several authors \cite{Contaldi:2003zv}. As mentioned before, this
can be naturally explained in our model, due to the suppression
behavior of $|\mathcal{A}(\vk)|$, $g(k) \sim -1/k^2$, (see Figs.
\ref{fig_leading} and \ref{fig_suppresion}).
%\begin{figure}[h]
%    \begin{minipage}{0.4\textwidth}
%    \includegraphics[width=7cm]{deformation.eps}
%    \caption{Deformation factor $|\mathcal{A}(\vk)|$ (for a given direction $\hat{\vk}$). Here we have chosen that $\vk$ is parrallel to $\va$ without generality. Parameters are chosen as $A=10^{-3}$, $c=2, n_s=0.96, \epsilon=0.01$. We plot $k$ from $10^{-5}$ to $10^{-2}$.}
%    \label{fig_deformation}
%    \end{minipage}
%\end{figure}

\section{Conclusion and Discussion}

In this work, we studied cosmological perturbations in the presence
of a superhorizon inhomogeneous inflaton background. We find that
the leading-order correction to the standard isotropic power
spectrum is a quadrupole anisotropy. Moreover, the deviation from
isotropy is generally scale-dependent. More specifically, the
quadrupole correction term has a suppression factor $g(k) \sim
-1/k^2$. And thus our model can simultaneously explain the observed
anisotropic alignment of the low-$\ell$ multipoles and their low
power. If the indication of these large scale CMB anomalies were to
be confirmed by future experiments, the work presented here would
improve our understanding of inflation and the very early universe.

In our work, the inhomogeneous background affects the perturbation
through the non-linear structure of $P(X,\phi)$ of $X$. A similar
effect would arise when one consider backreaction of superhorizon
perturbations through the nonlinearity of the Einstein equations
\cite{Mukhanov:1996ak}.

In our approximations, we neglected the backreaction of the
inhomogeneous inflaton background to the geometry, which has to be
considered in a more rigorous analysis. In that case, a non-FRW
metric may arise. For instance, in \cite{Gumrukcuoglu:2007bx},
anisotropic inflationary models and perturbations are investigated,
where an anisotropic spectrum is also produced. In a FRW background,
the quantization procedure is well understood for canonical variables
on sub-Hubble scales, where they evolve adiabatically, and thus the
adiabatic vacuum is well defined. In this work, we choose the mode
solution (\ref{mode_solution}) by simply noticing that
(\ref{mode_solution}) reduces to the standard mode solution in the
$A_{i}\rightarrow 0$ limit. It should be noted, however, that the
frequency term in the second-line of (\ref{mode_eom}) is divergent
when $\eta \rightarrow -\infty$, due to the presence of the
anisotropic term $\gamma (\va\cdot \vk)^2/(aH)^2$. Thus, the
presence of a WKB regime and also the quantization procedure in this
case are non-trivial in general (see e.g. discussions in
\cite{Gumrukcuoglu:2007bx}). We would like to investigate these
problems in future work.

Indeed, the work presented here is a general framework, while the
results of our model are generic. A more detailed study including
metric perturbations and possible non-FRW background geometry is
needed. We would like to present it in a companion work
\cite{me:wip}.

%%%%%%%%%%%%%%%%%%%%%%%%%%%%%%%%%%%%%%%%%%%%%%%%%%%%%%%%%%%%%%%%%%%%%%%%%%%%%%%%

\acknowledgements

I would like to thank Robert Brandenberger, Miao Li, Tao Wang, Yi
Wang, Bin Hu, for useful discussions and comments. I am grateful to
Robert Brandenberger for a careful reading the manuscript. This work
was supported by Grants of NSFC.

%%%%%%%%%%%%%%%%%%%%%%%%%%%%%%%%%%%%%%%%%%%%%%%%%%%%%%%%%
%\appendix
%
%\section{Effective Mass Term}
%The effective mass term is \ea{
%    \mathcal{M} &= \frac{P_{,\phi\phi}}{\Sigma^{00}} +
%    \frac{1}{a^2}\left[ 2\alpha \mathcal{H} - \alpha^2 + \alpha' +(c_s^2)_{ij}\beta_i\beta_j \right. \\
%    &\qquad +\lrp{\frac{\sqrt{\epsilon}\gamma}{aH}A_i\beta_i}'  + 2(\mathcal{H}-\alpha)\lrp{ \frac{\sqrt{\epsilon}\gamma}{aH}A_i\beta_i - \frac{aP_{,X\phi}\dot{\bap}}{\Sigma^{00}} }\\
%    &\qquad \qquad \left. -2\lrp{ \frac{aP_{,X\phi}\dot{\bap}}{\Sigma^{00}} }'  - \frac{2P_{,X\phi}}{\Sigma^{00}}A_i\beta_i
%    \right] \,,
%} where \ea{
%    \alpha &= \sqrt{-\Sigma^{00}} \rd_{\eta}
%    \frac{1}{\sqrt{-\Sigma^{00}}} \;,\\
%    \beta_i &= \sqrt{-\Sigma^{00}} \rd_{i}
%    \frac{1}{\sqrt{-\Sigma^{00}}} \;,\\
%    (c_s^2)_{ij} &= c^2\delta_{ij} + \frac{\gamma A_iA_j}{a^2H^2} \;.
%} In this work, we assume $\beta_i \approx 0$.

%%********************  Appendix.End  ********************%%

%%********************  Bibliography.Begin  ********************%%

%%********************  Bibliography.End  ********************%%


\begin{thebibliography}{99}

%%%%%%%%%%%%%%%%%%%%%%%%%%%%%%%%%%%%%%%%%%%%%%%%
%%  WMAP5

\bibitem{Komatsu:2008hk}
  E.~Komatsu {\it et al.}  [WMAP Collaboration],
  %``Five-Year Wilkinson Microwave Anisotropy Probe (WMAP\altaffilmark 1 )
  %Observations:Cosmological Interpretation,''
  Astrophys.\ J.\ Suppl.\  {\bf 180}, 330 (2009)
  %[arXiv:0803.0547 [astro-ph]].
  %%CITATION = APJSA,180,330;%%




%%%%%%%%%%%%%%%%%%%%%%%%%%%%%%%%%%%%%%%%%%%%%%%%%%%%%%%%%%%%%%%%%
%%  other theoretical analysis

%\bibitem{Cresswell:2005sh}
%  J.~G.~Cresswell, A.~R.~Liddle, P.~Mukherjee and A.~Riazuelo,
%  %``Cosmic microwave background multipole alignments in slab topologies,''
%  Phys.\ Rev.\  D {\bf 73}, 041302 (2006)
%  [arXiv:astro-ph/0512017].
%\bibitem{Alexander:2006mt}
%  S.~H.~S.~Alexander,
%  %``Is cosmic parity violation responsible for the anomalies in the WMAP
%  %data?,''
%  Phys.\ Lett.\  B {\bf 660}, 444 (2008)
%  [arXiv:hep-th/0601034].
%\bibitem{Berera:2003tf}
%  A.~Berera, R.~V.~Buniy and T.~W.~Kephart,
%  %``The eccentric universe,''
%  JCAP {\bf 0410}, 016 (2004)
%  [arXiv:hep-ph/0311233].
\bibitem{Pullen:2007tu}
  A.~R.~Pullen and M.~Kamionkowski,
  %``Cosmic Microwave Background Statistics for a Direction-Dependent Primordial
  %Power Spectrum,''
  Phys.\ Rev.\  D {\bf 76}, 103529 (2007)
  %[arXiv:0709.1144 [astro-ph]].
%\bibitem{Ando:2008zza}
  S.~Ando and M.~Kamionkowski,
  %``Nonlinear Evolution of Anisotropic Cosmological Power,''
  Phys.\ Rev.\ Lett.\  {\bf 100}, 071301 (2008)
  %[arXiv:0711.0779 [astro-ph]].
%%\bibitem{Buniy:2005qm}
%  R.~V.~Buniy, A.~Berera and T.~W.~Kephart,
%  %``Asymmetric inflation: Exact solutions,''
%  Phys.\ Rev.\  D {\bf 73}, 063529 (2006)
%  %[arXiv:hep-th/0511115].
%\bibitem{Chibisov:1990bk}
  G.~V.~Chibisov and Yu.~V.~Shtanov,
  %``Chaotic inflationary universe and the anisotropy of the large scale
  %structure,''
  Int.\ J.\ Mod.\ Phys.\  A {\bf 5} (1990) 2625.
%\bibitem{Chibisov:1989wb}
  G.~V.~Chibisov and Yu.~V.~Shtanov,
  %``STRUCTURAL ANISOTROPY IN A CHAOTIC INFLATIONARY UNIVERSE,''
  Sov.\ Phys.\ JETP {\bf 69} (1989) 17
  [Zh.\ Eksp.\ Teor.\ Fiz.\  {\bf 96} (1989) 32].
%\bibitem{Battye:2006mb}
  R.~A.~Battye and A.~Moss,
  %``Anisotropic perturbations due to dark energy,''
  Phys.\ Rev.\  D {\bf 74}, 041301 (2006)
  %[arXiv:astro-ph/0602377].
%\bibitem{Lerner:2008ad}
  R.~Lerner and J.~McDonald,
  %``Space-Dependent Step Features: Transient Breakdown of Slow-roll,
  %Homogeneity and Isotropy During Inflation,''
  Phys.\ Rev.\  D {\bf 79}, 023511 (2009)
  %[arXiv:0811.1933 [astro-ph]].
  %%CITATION = PHRVA,D79,023511;%%
%\bibitem{Dvorkin:2007jp}
  C.~Dvorkin, H.~V.~Peiris and W.~Hu,
  %``Testable polarization predictions for models of CMB isotropy anomalies,''
  Phys.\ Rev.\  D {\bf 77}, 063008 (2008)
  %[arXiv:0711.2321 [astro-ph]].
  %%CITATION = PHRVA,D77,063008;%%
%\bibitem{Moffat:2005yx}
  J.~W.~Moffat,
  %``Cosmic Microwave Background, Accelerating Universe and Inhomogeneous
  %Cosmology,''
  JCAP {\bf 0510}, 012 (2005)
  %[arXiv:astro-ph/0502110].
  %%CITATION = JCAPA,0510,012;%%
%\bibitem{Gordon:2006ag}
  C.~Gordon,
  %``Broken Isotropy from a Linear Modulation of the Primordial Perturbations,''
  Astrophys.\ J.\  {\bf 656}, 636 (2007)
  %[arXiv:astro-ph/0607423].
  %%CITATION = ASJOA,656,636;%%
%\bibitem{Akofor:2007fv}
  E.~Akofor, A.~P.~Balachandran, S.~G.~Jo, A.~Joseph and B.~A.~Qureshi,
  %``Direction-Dependent CMB Power Spectrum and Statistical Anisotropy from
  %Noncommutative Geometry,''
  JHEP {\bf 0805}, 092 (2008)
  %[arXiv:0710.5897 [astro-ph]].
  %%CITATION = JHEPA,0805,092;%%
%\bibitem{Copi:2008hw}
  C.~J.~Copi, D.~Huterer, D.~J.~Schwarz and G.~D.~Starkman,
  %``No large-angle correlations on the non-Galactic microwave sky,''
  %arXiv:0808.3767 [astro-ph].
  %%CITATION = ARXIV:0808.3767;%%
%\bibitem{Copi:2003kt}
  C.~J.~Copi, D.~Huterer and G.~D.~Starkman,
  %``Multipole Vectors--a new representation of the CMB sky and evidence for
  %statistical anisotropy or non-Gaussianity at 2<=l<=8,''
  Phys.\ Rev.\  D {\bf 70}, 043515 (2004)
  %[arXiv:astro-ph/0310511].
  %%CITATION = PHRVA,D70,043515;%%


%%%%%%%%%%%%%%%%%%%%%%%%%%%%%%%%%%%%%
%%  Suppression low l -- theory

\bibitem{Contaldi:2003zv}
  C.~R.~Contaldi, M.~Peloso, L.~Kofman and A.~Linde,
  %``Suppressing the lower Multipoles in the CMB Anisotropies,''
  JCAP {\bf 0307}, 002 (2003)
  %[arXiv:astro-ph/0303636].
  %%CITATION = JCAPA,0307,002;%%
%\bibitem{Nicholson:2007by}
  G.~Nicholson and C.~R.~Contaldi,
  %``The large scale CMB cut-off and the tensor-to-scalar ratio,''
  JCAP {\bf 0801}, 002 (2008)
  %[arXiv:astro-ph/0701783].
  %%CITATION = JCAPA,0801,002;%%
%\bibitem{de OliveiraCosta:2003pu}
  A.~de Oliveira-Costa, M.~Tegmark, M.~Zaldarriaga and A.~Hamilton,
  %``The significance of the largest scale CMB fluctuations in WMAP,''
  Phys.\ Rev.\  D {\bf 69}, 063516 (2004)
  %[arXiv:astro-ph/0307282].
  %%CITATION = PHRVA,D69,063516;%%
%\bibitem{Cline:2003ve}
  J.~M.~Cline, P.~Crotty and J.~Lesgourgues,
  %``Does the small CMB quadrupole moment suggest new physics?,''
  JCAP {\bf 0309}, 010 (2003)
  %[arXiv:astro-ph/0304558].
  %%CITATION = JCAPA,0309,010;%%
%\bibitem{Boyanovsky:2006qi}
  D.~Boyanovsky, H.~J.~de Vega and N.~G.~Sanchez,
  %``CMB quadrupole suppression: I. Initial conditions of inflationary
  %perturbations,''
  Phys.\ Rev.\  D {\bf 74}, 123006 (2006)
  %[arXiv:astro-ph/0607508].
  %%CITATION = PHRVA,D74,123006;%%
%\bibitem{Land:2005cg}
  K.~Land and J.~Magueijo,
  %``Template fitting and the large-angle CMB anomalies,''
  Mon.\ Not.\ Roy.\ Astron.\ Soc.\  {\bf 367}, 1714 (2006)
  %[arXiv:astro-ph/0509752].
  %%CITATION = MNRAA,367,1714;%%
%\bibitem{Sinha:2005mn}
  R.~Sinha and T.~Souradeep,
  %``Post-WMAP assessment of infrared cutoff in the primordial spectrum from
  %inflation,''
  Phys.\ Rev.\  D {\bf 74}, 043518 (2006)
  %[arXiv:astro-ph/0511808].
  %%CITATION = PHRVA,D74,043518;%%


%%%%%%%%%%%%%%%%%%%%%%%%%%%%%%%%%%%%%%%%%%%%%%%
%%  ACW

\bibitem{Ackerman:2007nb}
  L.~Ackerman, S.~M.~Carroll and M.~B.~Wise,
  %``Imprints of a Primordial Preferred Direction on the Microwave Background,''
  Phys.\ Rev.\  D {\bf 75}, 083502 (2007)
  %[arXiv:astro-ph/0701357].

%%%%%%%%%%%%%%%%%%%%%%%%%%%%%%%%%%%%%%%%%%%%%%%%
%%  axis of evil

\bibitem{Land:2005ad}
  K.~Land and J.~Magueijo,
  %``The axis of evil,''
  Phys.\ Rev.\ Lett.\  {\bf 95}, 071301 (2005)
  %[arXiv:astro-ph/0502237].
  %%CITATION = PRLTA,95,071301;%%
%\bibitem{de OliveiraCosta:2003pu}
  A.~de Oliveira-Costa, M.~Tegmark, M.~Zaldarriaga and A.~Hamilton,
  %``The significance of the largest scale CMB fluctuations in WMAP,''
  Phys.\ Rev.\  D {\bf 69}, 063516 (2004)
  %[arXiv:astro-ph/0307282].
  %%CITATION = PHRVA,D69,063516;%%

%%%%%%%%%%%%%%%%%%%%%%%%%%%%%%%%%%%%%%%%%%%%%%%%%%%%%%
%%  single mode

\bibitem{Gordon:2005ai}
  C.~Gordon, W.~Hu, D.~Huterer and T.~M.~Crawford,
  %``Spontaneous Isotropy Breaking: A Mechanism for CMB Multipole Alignments,''
  Phys.\ Rev.\  D {\bf 72}, 103002 (2005)
  %[arXiv:astro-ph/0509301].

\bibitem{Erickcek:2008sm}
  A.~L.~Erickcek, M.~Kamionkowski and S.~M.~Carroll,
  %``A Hemispherical Power Asymmetry from Inflation,''
  Phys.\ Rev.\  D {\bf 78}, 123520 (2008)
  %[arXiv:0806.0377 [astro-ph]].
  %%CITATION = PHRVA,D78,123520;%%
%\bibitem{Erickcek:2008jp}
  A.~L.~Erickcek, S.~M.~Carroll and M.~Kamionkowski,
  %``Superhorizon Perturbations and the Cosmic Microwave Background,''
  Phys.\ Rev.\  D {\bf 78}, 083012 (2008)
  %[arXiv:0808.1570 [astro-ph]].
  %%CITATION = PHRVA,D78,083012;%%





%%%%%%%%%%%%%%%%%%%%%%%%%%%%%%%%%%%%%%%%%%%%%%%%%%
%%  observation

\bibitem{ArmendarizPicon:2008yr}
  C.~Armendariz-Picon and L.~Pekowsky,
  %``Bayesian Limits on Primordial Isotropy Breaking,''
  Phys.\ Rev.\ Lett.\  {\bf 102}, 031301 (2009)
  %[arXiv:0807.2687 [astro-ph]].
  %%CITATION = PRLTA,102,031301;%%
%\bibitem{Hoftuft:2009rq}
  J.~Hoftuft, H.~K.~Eriksen, A.~J.~Banday, K.~M.~Gorski, F.~K.~Hansen and P.~B.~Lilje,
  %``Increasing evidence for hemispherical power asymmetry in the five-year WMAP
  %data,''
  arXiv:0903.1229 [astro-ph.CO].
  %%CITATION = ARXIV:0903.1229;%%
%\bibitem{Hansen:2008ym}
  F.~K.~Hansen, A.~J.~Banday, K.~M.~Gorski, H.~K.~Eriksen and P.~B.~Lilje,
  %``Power Asymmetry in Cosmic Microwave Background Fluctuations from Full Sky
  %to Sub-degree Scales: Is the Universe Isotropic?,''
  arXiv:0812.3795 [astro-ph].
  %%CITATION = ARXIV:0812.3795;%%
%\bibitem{Samal:2008nv}
  P.~K.~Samal, R.~Saha, P.~Jain and J.~P.~Ralston,
  %``Signals of Statistical Anisotropy in WMAP Foreground-Cleaned Maps,''
  arXiv:0811.1639 [astro-ph].
  %%CITATION = ARXIV:0811.1639;%%
%\bibitem{Eriksen:2007pc}
  H.~K.~Eriksen, A.~J.~Banday, K.~M.~Gorski, F.~K.~Hansen and P.~B.~Lilje,
  %``Hemispherical power asymmetry in the three-year Wilkinson Microwave
  %Anisotropy Probe sky maps,''
  Astrophys.\ J.\  {\bf 660}, L81 (2007)
  %[arXiv:astro-ph/0701089].
  %%CITATION = ASJOA,660,L81;%%
%\bibitem{Lew:2008mq}
  B.~Lew,
  %``Hemispherical power asymmetry: parameter estimation from CMB WMAP5 data,''
  JCAP {\bf 0809}, 023 (2008)
  %[arXiv:0808.2867 [astro-ph]].
  %%CITATION = JCAPA,0809,023;%%
%\bibitem{Groeneboom:2008fz}
  N.~E.~Groeneboom and H.~K.~Eriksen,
  %``Bayesian analysis of sparse anisotropic universe models and application to
  %the 5-yr WMAP data,''
  Astrophys.\ J.\  {\bf 690}, 1807 (2009)
  %[arXiv:0807.2242 [astro-ph]].
  %%CITATION = ASJOA,690,1807;%%
%\bibitem{Copi:2006tu}
  C.~Copi, D.~Huterer, D.~Schwarz and G.~Starkman,
  %``The Uncorrelated Universe: Statistical Anisotropy and the Vanishing Angular
  %Correlation Function in WMAP Years 1-3,''
  Phys.\ Rev.\  D {\bf 75}, 023507 (2007)
  %[arXiv:astro-ph/0605135].
  %%CITATION = PHRVA,D75,023507;%%
%%\bibitem{Eriksen:2003db}
%  H.~K.~Eriksen, F.~K.~Hansen, A.~J.~Banday, K.~M.~Gorski and P.~B.~Lilje,
%  %``Asymmetries in the CMB anisotropy field,''
%  Astrophys.\ J.\  {\bf 605}, 14 (2004)
%  [Erratum-ibid.\  {\bf 609}, 1198 (2004)]
%  %[arXiv:astro-ph/0307507].
%  %%CITATION = ASJOA,605,14;%%
%%\bibitem{Hajian:2004zn}
%  A.~Hajian, T.~Souradeep and N.~J.~Cornish,
%  %``Statistical Isotropy of the WMAP Data: A Bipolar Power Spectrum Analysis,''
%  Astrophys.\ J.\  {\bf 618}, L63 (2004)
%  %[arXiv:astro-ph/0406354].
%  %%CITATION = ASJOA,618,L63;%%
%%\bibitem{Hajian:2003qq}
%  A.~Hajian and T.~Souradeep,
%  %``Measuring Statistical isotropy of the CMB anisotropy,''
%  Astrophys.\ J.\  {\bf 597}, L5 (2003)
%  %[arXiv:astro-ph/0308001].
%  %%CITATION = ASJOA,597,L5;%%
%\bibitem{Kahniashvili:2008va}
  T.~Kahniashvili, R.~Durrer and Y.~Maravin,
  %``Testing Lorentz Invariance Violation with WMAP Five Year Data,''
  Phys.\ Rev.\  D {\bf 78}, 123009 (2008)
  %[arXiv:0807.2593 [astro-ph]].
  %%CITATION = PHRVA,D78,123009;%%
%\bibitem{Hajian:2006ud}
  A.~Hajian and T.~Souradeep,
  %``Testing Global Isotropy of Three-Year Wilkinson Microwave Anisotropy Probe
  %(WMAP) Data: Temperature Analysis,''
  Phys.\ Rev.\  D {\bf 74}, 123521 (2006)
  %[arXiv:astro-ph/0607153].
  %%CITATION = PHRVA,D74,123521;%%
%%\bibitem{Hansen:2004mj}
%  F.~K.~Hansen, P.~Cabella, D.~Marinucci and N.~Vittorio,
%  %``Asymmetries in the local curvature of the WMAP data,''
%  Astrophys.\ J.\  {\bf 607}, L67 (2004)
%  %[arXiv:astro-ph/0402396].
%  %%CITATION = ASJOA,607,L67;%%
%%\bibitem{Hansen:2004vq}
%  F.~K.~Hansen, A.~J.~Banday and K.~M.~Gorski,
%  %``Testing the cosmological principle of isotropy: local power spectrum
%  %estimates of the WMAP data,''
%  arXiv:astro-ph/0404206.
%  %%CITATION = ASTRO-PH/0404206;%%
%%\bibitem{Covi:2006ci}
%  L.~Covi, J.~Hamann, A.~Melchiorri, A.~Slosar and I.~Sorbera,
%  %``Inflation and WMAP three year data: Features have a future!,''
%  Phys.\ Rev.\  D {\bf 74}, 083509 (2006)
%  %[arXiv:astro-ph/0606452].
%  %%CITATION = PHRVA,D74,083509;%%


%%%%%%%%%%%%%%%%%%%%%%%%%%%%%%%%%%%%%%%%%%%%%%
%%  initial conditions

\bibitem{Goldwirth:1991rj}
  D.~S.~Goldwirth and T.~Piran,
  %``Initial conditions for inflation,''
  Phys.\ Rept.\  {\bf 214}, 223 (1992).
  %%CITATION = PRPLC,214,223;%%
%\bibitem{Linde:1985ub}
  A.~D.~Linde,
  %``Initial Conditions For Inflation,''
  Phys.\ Lett.\  B {\bf 162} (1985) 281.
  %%CITATION = PHLTA,B162,281;%%
%\bibitem{Goldwirth:1989pr}
  D.~S.~Goldwirth and T.~Piran,
  %``INHOMOGENEITY AND THE ONSET OF INFLATION,''
  Phys.\ Rev.\ Lett.\  {\bf 64}, 2852 (1990).
  %%CITATION = PRLTA,64,2852;%%




%%%%%%%%%%%%%%%%%%%%%%%%%%%%%%%%%%%%%%%%%%%%%%%%%
%%  B's backreaction

\bibitem{Mukhanov:1996ak}
  V.~F.~Mukhanov, L.~R.~W.~Abramo and R.~H.~Brandenberger,
  %``On the back reaction problem for gravitational perturbations,''
  Phys.\ Rev.\ Lett.\  {\bf 78}, 1624 (1997)
  %[arXiv:gr-qc/9609026].
  %%CITATION = PRLTA,78,1624;%%
%\bibitem{Abramo:1996gd}
  L.~R.~W.~Abramo, R.~H.~Brandenberger and V.~F.~M.~Mukhanov,
  %``The back reaction of gravitational perturbations,''
  [arXiv:gr-qc/9702004].
  %%CITATION = GR-QC/9702004;%%
%\bibitem{Martineau:2005aa}
  P.~Martineau and R.~H.~Brandenberger,
  %``The effects of gravitational back-reaction on cosmological
  %perturbations,''
  Phys.\ Rev.\  D {\bf 72}, 023507 (2005)
  %[arXiv:astro-ph/0505236].
  %%CITATION = PHRVA,D72,023507;%%
%\bibitem{Martineau:2005zu}
  P.~Martineau and R.~Brandenberger,
  %``Back-reaction: A cosmological panacea,''
  [arXiv:astro-ph/0510523].
  %%CITATION = ASTRO-PH/0510523;%%



%%%%%%%%%%%%%%%%%%%%%%%%%%%%%
%%  gradient

\bibitem{ArmendarizPicon:2007nr}
  C.~Armendariz-Picon,
  %``Creating Statistically Anisotropic and Inhomogeneous Perturbations,''
  JCAP {\bf 0709}, 014 (2007)
  %[arXiv:0705.1167 [astro-ph]].
  %%CITATION = JCAPA,0709,014;%%

\bibitem{Donoghue:2007ze}
  J.~F.~Donoghue, K.~Dutta and A.~Ross,
  %``Non-isotropy in the CMB power spectrum in single field inflation,''
  [arXiv:astro-ph/0703455].
%\bibitem{Donoghue:2004gu}
  E.~P.~Donoghue and J.~F.~Donoghue,
  %``Isotropy of the early universe from CMB anisotropies,''
  Phys.\ Rev.\  D {\bf 71}, 043002 (2005)
  %[arXiv:astro-ph/0411237].
  %%CITATION = PHRVA,D71,043002;%%

%%%%%%%%%%%%%%%%%%%%%%%%%%%%%%%%%%%%%%%%%%%%%%%%%%%%%%

\bibitem{me:wip}
    Xian Gao, work in progress.



%%%%%%%%%%%%%%%%%%%%%%%%%%%%%%%%%%%%%%
%%  vector
\bibitem{Yokoyama:2008xw}
  S.~Yokoyama and J.~Soda,
  %``Primordial statistical anisotropy generated at the end of inflation,''
  JCAP {\bf 0808}, 005 (2008)
  %[arXiv:0805.4265 [astro-ph]].
  %%CITATION = JCAPA,0808,005;%%
%\bibitem{Kanno:2008gn}
  S.~Kanno, M.~Kimura, J.~Soda and S.~Yokoyama,
  %``Anisotropic Inflation from Vector Impurity,''
  JCAP {\bf 0808}, 034 (2008)
  [arXiv:0806.2422 [hep-ph]].
  %%CITATION = JCAPA,0808,034;%%
%\bibitem{Watanabe:2009ct}
  M.~a.~Watanabe, S.~Kanno and J.~Soda,
  %``Hairy Inflation,''
  arXiv:0902.2833 [hep-th].
  %%CITATION = ARXIV:0902.2833;%%
%\bibitem{Dimopoulos:2008yv}
  K.~Dimopoulos, D.~H.~Lyth and Y.~Rodriguez,
  %``Statistical anisotropy of the curvature perturbation from vector field
  %perturbations,''
  arXiv:0809.1055 [astro-ph].
  %%CITATION = ARXIV:0809.1055;%%
%%\bibitem{Karciauskas:2008bc}
%  M.~Karciauskas, K.~Dimopoulos and D.~H.~Lyth,
%  %``Anisotropic non-Gaussianity from vector field perturbations,''
%  arXiv:0812.0264 [astro-ph].
%  %%CITATION = ARXIV:0812.0264;%%
%\bibitem{Kahniashvili:2008sh}
  T.~Kahniashvili, G.~Lavrelashvili and B.~Ratra,
  %``CMB Temperature Anisotropy from Broken Spatial Isotropy due to an
  %Homogeneous Cosmological Magnetic Field,''
  Phys.\ Rev.\  D {\bf 78}, 063012 (2008)
  %[arXiv:0807.4239 [astro-ph]].
  %%CITATION = PHRVA,D78,063012;%%

%%%%%%%%%%%%%%%%%%%%%%%%%%%%%%%%%%%%
%%  Bianchi
\bibitem{Gumrukcuoglu:2007bx}
  A.~E.~Gumrukcuoglu, C.~R.~Contaldi and M.~Peloso,
  %``Inflationary perturbations in anisotropic backgrounds and their imprint on
  %the CMB,''
  JCAP {\bf 0711}, 005 (2007)
  %[arXiv:0707.4179 [astro-ph]].
%%\bibitem{Gumrukcuoglu:2006xj}
%  A.~E.~Gumrukcuoglu, C.~R.~Contaldi and M.~Peloso,
%  %``CMB Anomalies from Relic Anisotropy,''
%  arXiv:astro-ph/0608405.
%  %%CITATION = ASTRO-PH/0608405;%%
%%\bibitem{Himmetoglu:2008zp}
%  B.~Himmetoglu, C.~R.~Contaldi and M.~Peloso,
%  %``Instability of anisotropic cosmological solutions supported by vector
%  %fields,''
%  arXiv:0809.2779 [astro-ph].
%  %%CITATION = ARXIV:0809.2779;%%
%\bibitem{Pitrou:2008gk}
  C.~Pitrou, T.~S.~Pereira and J.~P.~Uzan,
  %``Predictions from an anisotropic inflationary era,''
  JCAP {\bf 0804}, 004 (2008)
  %[arXiv:0801.3596 [astro-ph]].
  %%CITATION = JCAPA,0804,004;%%
%%\bibitem{Pereira:2007yy}
%  T.~S.~Pereira, C.~Pitrou and J.~P.~Uzan,
%  %``Theory of cosmological perturbations in an anisotropic universe,''
%  JCAP {\bf 0709}, 006 (2007)
%  %[arXiv:0707.0736 [astro-ph]].
%  %%CITATION = JCAPA,0709,006;%%
%\bibitem{Boehmer:2007ut}
  C.~G.~Boehmer and D.~F.~Mota,
  %``CMB Anisotropies and Inflation from Non-Standard Spinors,''
  Phys.\ Lett.\  B {\bf 663}, 168 (2008)
  %[arXiv:0710.2003 [astro-ph]].






%%%%%%%%%%%%%%%%%%%%%%%%%%%%%%%%%%%%%%%%%%%%%%%%%%%%%%%%%%%%%%%%%%%%%%%%%%%
%%%%%%%%%%%%%%%%%%%%%%%%%%%%%%%%%%%%%%%%%%%%%%%%%%%%%%%%%%%%%%%%%%%%%%%%%%%
%%%%  others

%%%%%%%%%%%%%%%%%%%%%%%%%%%%%%%%%%%%%%%%%%
%%  multipole vectors

%\bibitem{Copi:2003kt}
%  C.~J.~Copi, D.~Huterer and G.~D.~Starkman,
%  %``Multipole Vectors--a new representation of the CMB sky and evidence for
%  %statistical anisotropy or non-Gaussianity at 2<=l<=8,''
%  Phys.\ Rev.\  D {\bf 70}, 043515 (2004)
%  [arXiv:astro-ph/0310511].
%  %%CITATION = PHRVA,D70,043515;%%
%%\bibitem{Helling:2006xh}
%  R.~C.~Helling, P.~Schupp and T.~Tesileanu,
%  %``CMB statistical anisotropy, multipole vectors and the influence of the
%  %dipole,''
%  Phys.\ Rev.\  D {\bf 74}, 063004 (2006)
%  [arXiv:astro-ph/0603594].
%  %%CITATION = PHRVA,D74,063004;%%

%%%%%%%%%%%%%%%%%%%%%%%%%%%%%%%%%%%%%%%%%%
%%  other general formulation

%\bibitem{Pullen:2007tu}
%  A.~R.~Pullen and M.~Kamionkowski,
%  %``Cosmic Microwave Background Statistics for a Direction-Dependent Primordial
%  %Power Spectrum,''
%  Phys.\ Rev.\  D {\bf 76}, 103529 (2007)
%  [arXiv:0709.1144 [astro-ph]].
%%\bibitem{Ando:2008zza}
%  S.~Ando and M.~Kamionkowski,
%  %``Nonlinear Evolution of Anisotropic Cosmological Power,''
%  Phys.\ Rev.\ Lett.\  {\bf 100}, 071301 (2008)
%  [arXiv:0711.0779 [astro-ph]].
%%%\bibitem{ArmendarizPicon:2005jh}
%  C.~Armendariz-Picon,
%  %``Footprints of Statistical Anisotropies,''
%  JCAP {\bf 0603}, 002 (2006)
%  [arXiv:astro-ph/0509893].
%  %%CITATION = JCAPA,0603,002;%%


%%%%%%%%%%%%%%%%%%%%%%%%%%%%%%%%%%%%%%%%%
%%  topology

%\bibitem{Levin:2001fg}
%  J.~J.~Levin,
%  %``Topology and the cosmic microwave background,''
%  Phys.\ Rept.\  {\bf 365}, 251 (2002)
%  [arXiv:gr-qc/0108043].
%  %%CITATION = PRPLC,365,251;%%

\end{thebibliography}
\end{document}